\documentclass[a4paper]{jpconf}
\usepackage{graphicx}
\usepackage{iopams}
\begin{document}
\title{Cage multipoles in rare-earth hexaborides}

\author{M Amara$^{1,2}$, R M Gal\'era$^{1,2}$}
\address{$^1$ Univ. Grenoble Alpes, Inst NEEL, F-38000 Grenoble, France}
\address{$^2$ CNRS, Inst NEEL, F-38042 Grenoble, France}

\ead{mehdi.amara@neel.cnrs.fr}

\begin{abstract}
In rare-earth cage compounds, the guest 4$f$ ion cannot be considered as fixed at the centre of its cage. As result of the electronic degeneracy of the 4$f$ shell, single-ion or collective mechanisms can redistribute the ion inside the cage, which can be described in terms of multipolar components. These mechanisms and their influence are here discussed and illustrated in relation with the rare-earth hexaboride series. 
\end{abstract}

\section{Introduction}
Rare-earth (RE) cage compounds have been extensively investigated in the past decade, much interest being attracted by the filled skutterudites series \cite{Jeitschko1977}. These compounds display a variety of properties such as heavy fermion behaviour, unconventional, non-magnetic orderings, superconductivity etc. \cite{Bauer2002,Iwasa2005}. In the interpretation of their properties, little reference to their specific crystallography, that offers a considerable latitude of movement to the RE guest, can be found in the literature. A better understanding of the physical mechanisms active in these compounds could be inspired by an older, much investigated cage-type series, the RE hexaborides (RB$_6$).\\
In this series, specific phenomena develop that illustrate how displacements of the RE ions can lower the system's energy. The latitude of movement of the ion can be formalized using basic quantum mechanics, considering the 4$f$ ion trapped in an anharmonic potential that includes magnetic terms. The vibrational wave functions and energies determine a temperature dependent in-cage distribution of the RE and, for weakly separated levels, a high reactivity of this distribution to environmental perturbation. These perturbations can arise from pair interactions, as in the antiferromagnetic states of the RB$_6$ compounds. Then, concomitantly with the magnetic order, static displacement waves of the RE ions develop that reduce the RKKY exchange energy. Changes in the RE distribution inside the cage can also result from a single ion effect : in case of a high symmetry at the cage center, a centrifugal Jahn-Teller effect is likely to develop. This might shed light on some intriguing properties of CeB$_6$.\\ 

\section{The magnetic rattler}
Decoupling the vibrations of the 4$f$ ions is the most practical starting point for describing  "rattling" related properties. For the lowest vibration levels, the moving ion will induce little distortion of the lattice which can be considered as undeformable. From this local picture, couplings between the oscillators can be subsequently added, if required, at the expenses of the simplicity of the mathematical treatment.
 
\subsection{The simplest cage potential}
\begin{figure}[h]
\begin{minipage}{3in}
\includegraphics[trim = 0 0 0 5cm, clip, width=2.7in]{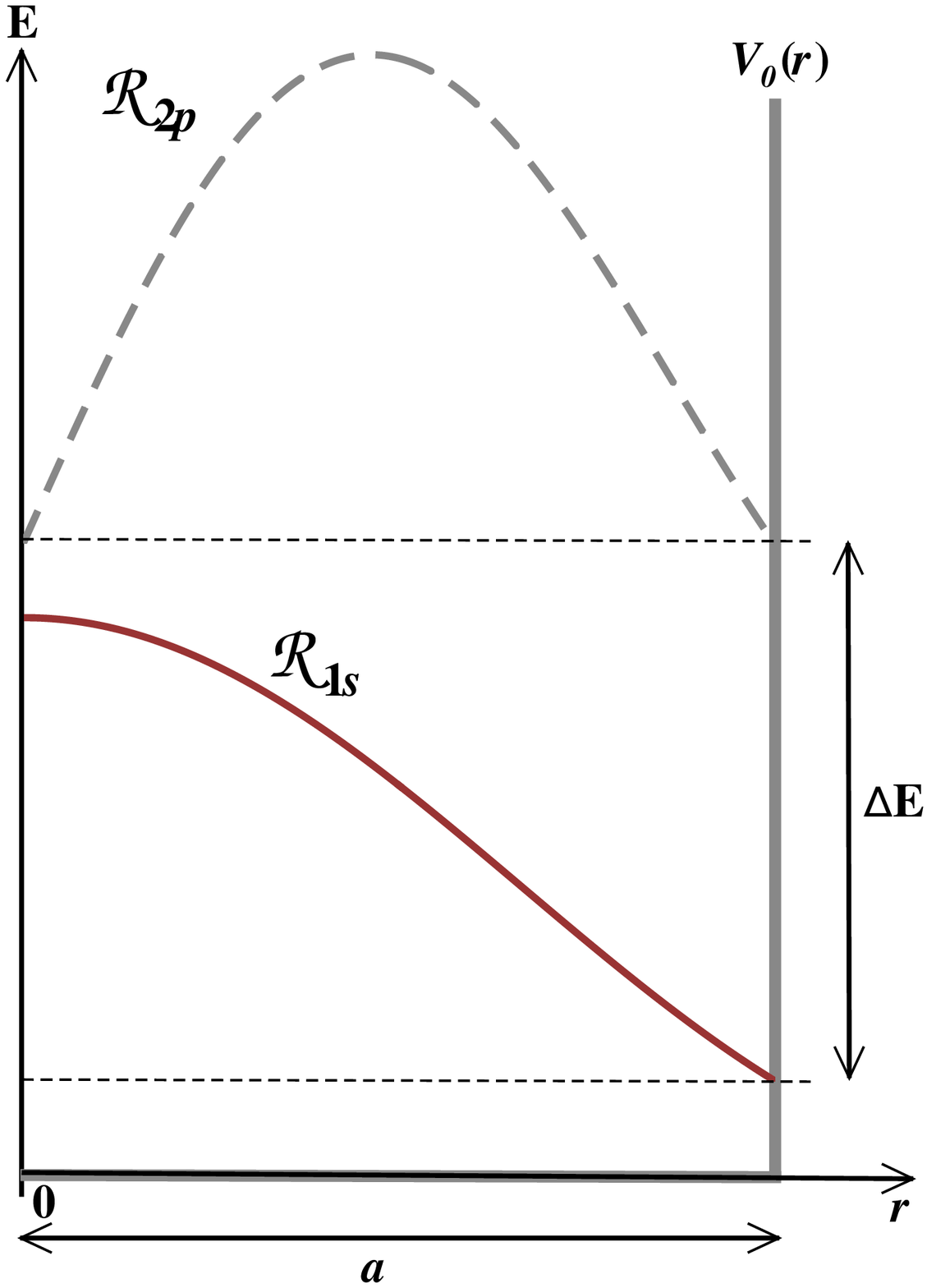}
\caption{\label{Fig1} The potential $V(r)$ of a spherical box of radius $a$ and sketches of the radial wave functions for the first two levels, $\mathcal{R}_{1s}$ and $\mathcal{R}_{2p}$, separated by an energy $\Delta E$.}
\end{minipage}\hspace{2pc}%
\begin{minipage}{3in}
\includegraphics[width=2.8in]{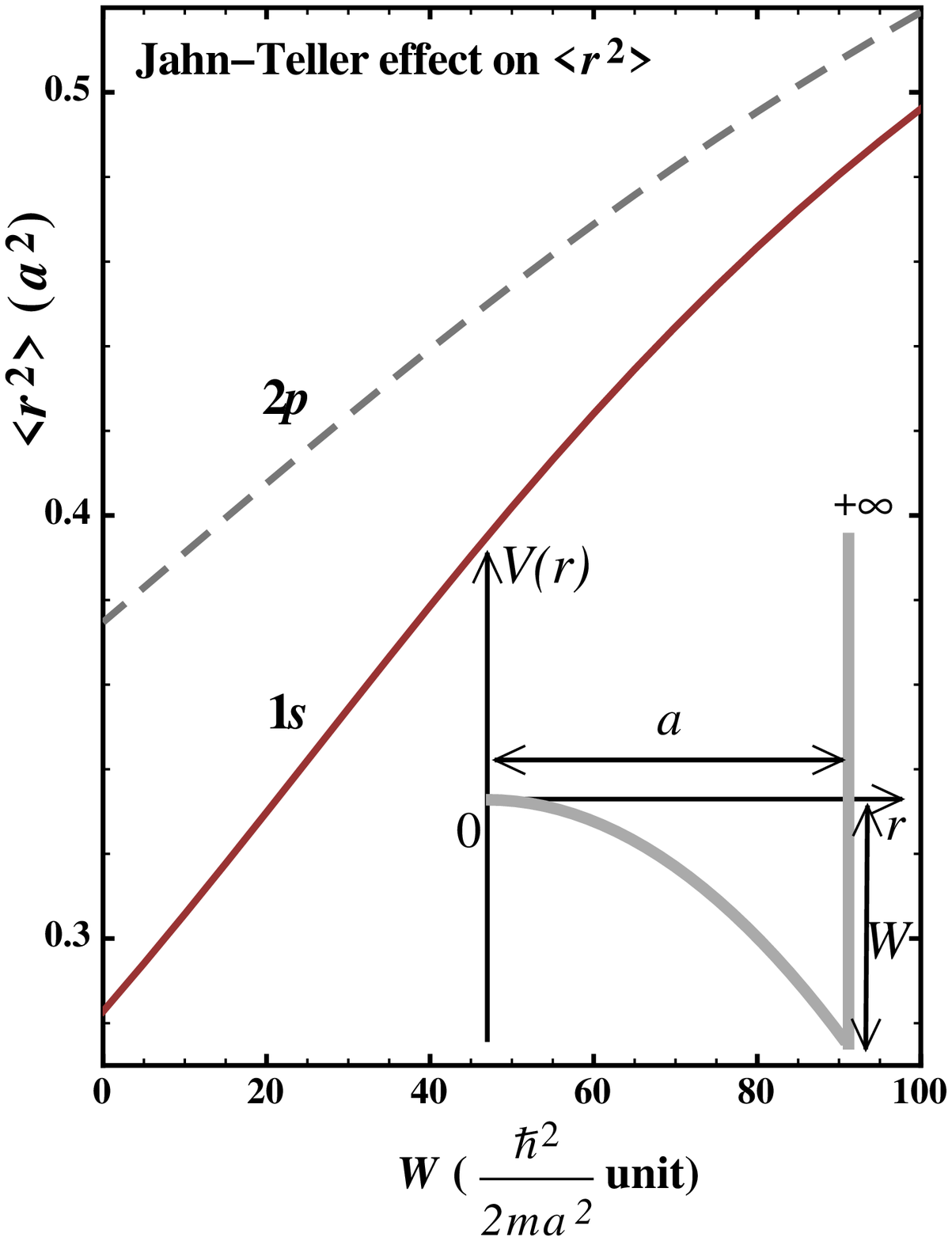}
\caption{\label{Fig2} Computed centrifugal Jahn-Teller effect on the 1$s$ and 2$p$ states, as reflected by the second moment $\langle r^2 \rangle$ of their radial distributions. The inset represents the cage potential, in the zero temperature limit, that includes a Jahn-Teller term of amplitude $W$.}
\end{minipage}
\end{figure}

For a RE ion in an oversized cage, an anharmonic potential should be considered. The simplest depiction is that of a flat potential that becomes abruptly infinite on the impenetrable cage walls ($V_{0}(r)$ on figure \ref{Fig1}). The energies and wave functions of the oscillator then depend entirely on the size $a$ of the cage and the mass $m$ of the ion. As regards the hexaborides, the excitation energy of the first vibration mode is found slightly larger than 10 meV (11 meV in PrB$_6$ \cite{Kohgi2006}, 12 meV in LaB$_6$ \cite{Smith1985}). This means that, at the low temperatures (T $<$ 50 K) where interesting phenomena take place, one can restrict to the ground and first excited vibrational levels. Then, the $O_h$ symmetry cage can be indifferently simplified to a spherical or cubic box. In both cases, the wave functions have identical transformation properties, the ground state being a singlet and the first excited a triplet. For convenience, we will adopt the spherical atomic like description, the first two vibration levels being then identified with the 1$s$ and 2$p$ states. Figure \ref{Fig1} includes sketches of their radial wave functions, obtained by numerically solving the time independent Schr\"{o}dinger equation. These two levels are separated by an energy $\Delta E = 10.3261(8) \frac{\hbar^2}{2ma^2}$. Relating this with the experimental separation of about 10 meV in the RB$_6$, one can estimate the radius of the model box at about 0.12 \AA . 

\subsection{Cage multipolar susceptibilities}
Starting from the simple spherical box above described, one can improve the description by adding terms in the potential, first treated as perturbations. Other corrections may stem from interactions of the encaged ion with its environment. These corrections will affect the distribution of the 4$f$ ion in the cage and, to evaluate their influence, it is convenient to invoke the multipolar susceptibilities of the system. In contrast with the 4$f$ shell terms, the weak separation between the 1$s$ and 2$p$ vibration levels allows odd cage multipoles to emerge in the RE ion's distribution. In particular, one can expect a significative dipolar susceptibility $\alpha_1$, which from perturbation theory, reads as:  
\begin{equation}
\label{suscept}
\alpha _1 (T) =  \frac{2}{3} \frac{\left|\langle \mathcal{R}_{1s} \right| r\left| \mathcal{R}_{2p}\rangle \right|^2 }{\Delta E} (1-\frac{2}{3+e^{\Delta E/k_{B}T}})
\end{equation}

\subsection{Crystal field contribution to the potential}
If the RE ion carries an orbital degeneracy, the Crystalline Electric Field (CEF) will split its electronic ground state. In the high symmetry at the cage centre, this splitting is only partial. However, if the 4$f$ ion departs from the centre, the symmetry of its environment is lowered and the CEF levels reach the minimal electrostatic degeneracy. Assuming that the electronic wave functions adapt rapidly to the new environment (Born-Oppenheimer), this can be translated into the interference, in the CEF hamiltonian of the 4$f$ ion, of a second order term:   
\begin{equation}
\label{HJT}
\mathcal{H}_{JT}(\bi{r})=-D^{\gamma} [(3 z^2-r^2) O_{2}^{0} + 3 (x^2-y^2) O_{2}^{2}]
-D^{\varepsilon} [(x \cdot y) P_{xy} + (y \cdot z) P_{yz} + (z \cdot x) P_{zx}]
\end{equation}
where $x$, $y$ and $z$ are the cartesian components of the displacement $\bi{r}$, along the cubic axes and $\{O_{2}^{0}$, $O_{2}^{2}\}$ and $\{P_{xy}$, $P_{yz}$, $P_{zx}\}$ are the 4$f$ quadrupolar operators \cite{MorinSchmitt1990}, which couple to the environment through the $D^{\gamma}$ and $D^{\varepsilon}$ constants. This CEF term cannot be neglected when describing the boxed ion : for a RE ion pressed against the cage, the splitting of the centre crystal field levels will be competitive with the weak separation of the 1$s$ and 2$p$ levels. This means that the 4$f$ electrons energy will significantly contribute to the cage potential. If the CEF ground state at the cage centre has an excess of orbital degeneracy, it determines a Jahn-Teller instability. At low temperature, the off-centre splitting of the ground state lowers the 4$f$ ion electrostatic energy, which results in a Jahn-Teller centrifugal force.\\
At zero temperature, this centrifugal term can be derived from equation \ref{HJT}, selecting the minimum CEF energy at any point $\bi{r}$ in the cage. With an additional spherical approximation, if $Q$ is the maximum absolute value of the induced quadrupole, the 4$f$ contribution to the cage potential reads as: 
\begin{equation}
\label{HJTQ}
V_{4f}(r)=-  \;D \; Q \; r^2
\end{equation}
where $D$ is the absolute value of the spherical quadrupolar coupling constant. The constant $W =\;D \; Q$ represents the strength of the Jahn-Teller centrifugal effect and introduces a new energy scale in the system. The total potential inside the cage becomes $V(r)=V_{0}(r)+V_{4f}(r)= - W r^2$ (see inset of figure \ref{Fig2}). The first-order correction associated with $V_{4f}(r)$, $E_{4f}^{1}=-  W\; \langle r^2\rangle$, directly reflects the centrifugal effect that favours an excentric distribution. In this regard, 2$p$ has an advantage over 1$s$ : the Jahn-Teller effect will decrease the energy separation between the two levels. Beyond a perturbative insight, this is illustrated by the dependence on $W$ of the numerically computed $\langle r^2\rangle$ (see figure \ref{Fig2}). The reduction of $\Delta E$ and the excentred distribution will both contribute to an increase of the odd multipolar susceptibilies (see equation \ref{suscept})\\
Another implication of equation \ref{HJT} is the spreading of the CEF levels. Even in absence of a Jahn-Teller effect, there is a widening of the CEF levels, due to the impossibility to perfectly localize the 4$f$ ion. The Jahn-Teller effect or, at finite temperature, the excited levels will increase the probability of presence of the ion at the cage periphery, further widening the CEF levels. The vibrations will tend to neutralize the CEF influence on the properties more rapidly, with increasing the temperature, than expected from considering the CEF scheme only.

\begin{table}[h]
\caption{\label{AF} Antiferromagnetic phase transitions in the RB$_6$ series.}
\begin{tabular}{lllll}
\br
 & T$_N$, T* (K) & wave vector & ordering process & References \\
\mr
CeB$_6$ &  2.4	&  $\langle1/4\; 1/4\; 1/2\rangle$,  $\langle1/4\; 1/4\; 0\rangle$ & 2$^{nd}$ order &\cite{Effantin1985}\\
PrB$_6$ & 6.9,  3.9* &  $\langle(1/4-\delta) \; 1/4\; 1/2\rangle$,  $\langle1/4\; 1/4\; 1/2\rangle$* & 1$^{st}$ order &\cite{McCarthy1980B}\\
NdB$_6$ & 8.6 &  $\langle1/2\; 0\; 0\rangle$ & 1$^{st}$ order & \cite{Geballe68}\\
GdB$_6$	& 16, 11* &  $\langle1/4\; 1/4\; 1/2\rangle$ & 1$^{st}$ order &\cite{Nozaki1980, Luca2004}\\
TbB$_6$	& 20 &  $\langle1/4\; 1/4\; 1/2\rangle$ & 1$^{st}$ order &\cite{Takahashi1997, Luca2004}\\
DyB$_6$	& 26 &  $\langle1/4\; 1/4\; 1/2\rangle$ & 1$^{st}$ order &\cite{Takahashi1997,Takahashi1998}\\
HoB$_6$	& 5.6 &  $\langle1/4\; 1/4\; 1/2\rangle$ & (?) &\cite{Segawa1992, Donni2001}\\
\br
\end{tabular}
\end{table}

\section{The displacive antiferromagnetism in the RB$_6$ compounds}
Most elements in the RB$_6$ series order antiferromagnetically (see table \ref{AF}). Remarquably, the singular magnetic wave vector $[1/4\; 1/4\; 1/2]$ is shared by most elements and, among them, many display a first-order magnetic ordering. This first-order process is particularly difficult to interpret for GdB$_6$ where no orbital effect can be invoked. X-ray experiments on GdB$_6$ revealed that, concomitantly with the magnetic order, static displacement waves of the Gd$^{3+}$ ions develop. Experiments on TbB$_6$ showed that these dipolar waves are not exclusive to GdB$_6$. In both cases, the displacements wave vectors correspond to the sum of two members from the magnetic star $\langle1/4\; 1/4\; 1/2\rangle$. Moreover, the temperature dependence of the displacement behaves as the square of the magnetic moment. The displacement is thus a secondary order parameter, induced by the primary, magnetic one. One has then to identify the force of magnetic origin that pushes the 4$f$ ions out of the cage centre. As the exchange interactions necessarily depend on the distance between the interacting 4$f$ ions, one can express the magnetic force $\bi{F}_i$ exerted on the ion at site $i$ \cite{Amara2005,Amara2010}. Taking account of the dipolar susceptibility $\alpha_1$, the statistical value of the displacement will write as :
\begin{equation}
\label{mag2disp}
\langle\bi{r}_{i}\rangle =\alpha_1 \langle \bi{F}_{i}\rangle=-\alpha_1 \sum\limits_{j,\,j\neq
i}(\bi{m}_{i} \cdot \bi{m}_{j})\, \bi{g}_{ij}
\end{equation}
where $\bi{m}_{i}$ is the statistical value of magnetic moment at site $i$ and $\bi{g}_{ij}$ is the gradient of the exchange coupling constant between sites $i$ and $j$. This explicits the quadratic relation between the displacement and the magnetism. Expressing the magnetic moment in terms of Fourier series, it follows that the displacements wave vectors $\bi q$ are such that $\bi q = \bi k_1 + \bi k_2$, where $\bi k_1$ and $\bi k_2$ belong to the $\langle1/4\; 1/4\; 1/2\rangle$ star. To minimize the exchange energy, the displacements corrected mean field has to gain amplitude at all sites, which imposes a condition on the magnetic wave vector \cite{Amara2010} : $\bi{k} = \bi{H}/4$, where $\bi{H}$ is a reciprocal lattice node. The $\langle1/4\; 1/4\; 1/2\rangle$ magnetic wave vector is one of the few that fulfill this condition, which explains its recurrence in the series. As regards the frequent first-order magnetic transition, it stems from the enhancement at $T_N$ of the exchange couplings with the emergence of the displacements.

\begin{figure}[h]
\begin{minipage}{3in}
\includegraphics[width=3in]{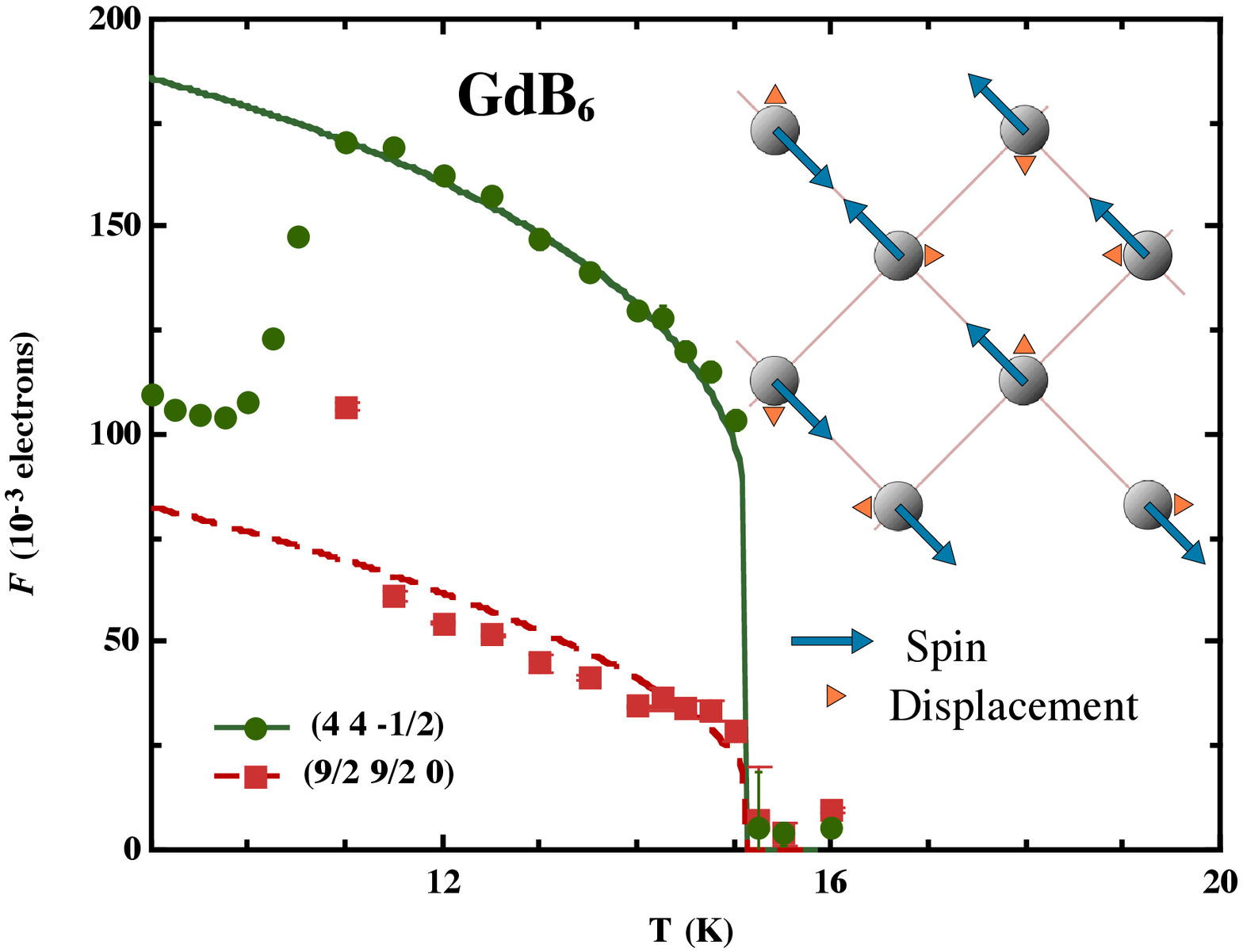}
\caption{\label{Fig3} Temperature dependence of the structure factors of representative X-ray reflections in GdB$_6$ (from \cite{Amara2005}). The lines are results from mean-field calculations and the inset shows sketches of the coexisting magnetic and displacive structures below T$_N$.}
\end{minipage}\hspace{2pc}%
\begin{minipage}{3in}
\includegraphics[trim = 0 0 0 0.3cm, clip, width=3in]{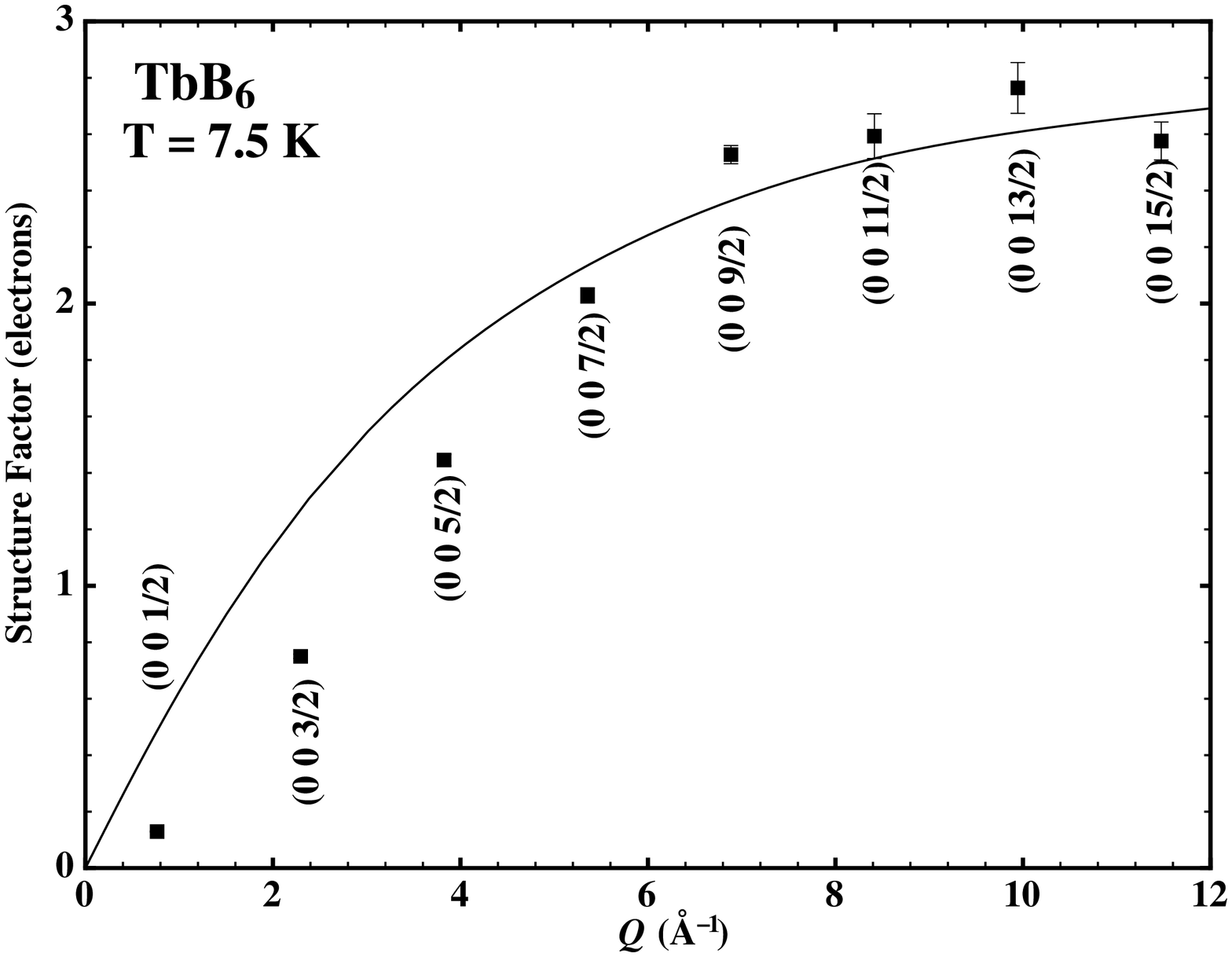}
\caption{\label{Fig4} Experimental structure factors of the specular $\langle00\frac{1}{2}\rangle$ -type reflections at 7.5 K in TbB$_6$ \cite{Amara2010}. The line represents the scaled  structure factor for a displacement wave of the Tb$^{3+}$ ions.}
\end{minipage}
\end{figure}

\section{Possible centrifugal Jahn-Teller effect in CeB$_6$}
\begin{figure}[h]
\begin{center}
\includegraphics[width=4in]{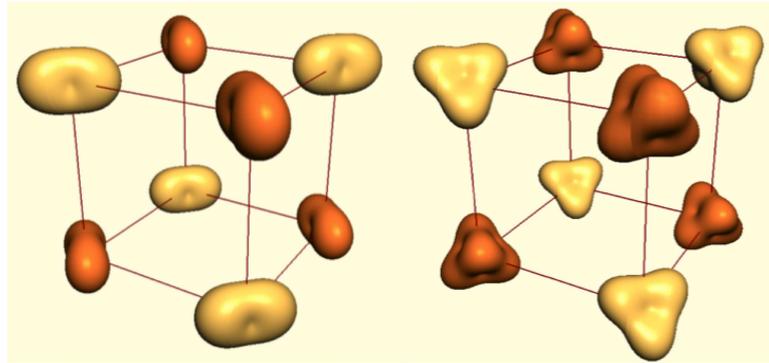}
\end{center}
\caption{\label{Fig5} Left, sketch of an antiferroquadrupolar, tetragonal, structure with $[1/2 \;1/2 \; 1/2]$ wave vector. Right, octupolar cubic structure with same wave vector, possibly realized by decentering the Ce$^{3+}$ in their cages. }
\end{figure}

The most investigated element in the RB$_6$ series is CeB$_6$. A large CEF splitting of the J = 5/2 multiplet, isolates a $\Gamma_8$ ground state quadruplet\cite{Zirngiebl1984}, 530 K below the $\Gamma_7$ doublet. This is twice the Kramers degeneracy and the conditions of a Jahn-Teller instability are realized : the quadruplet can split into two doublets, with the emergence of 4$f$ quadrupoles. As there exists in CeB$_6$ a non-magnetic ordering \cite{Effantin1985} below T$_Q$ = 3.2 K, associated with the [1/2~1/2~1/2] wave-vector \cite{Erkelens1987, Tanaka2004}, this state has been for years considered as an antiferroquadrupolar order : i.e., the 4$f$ quadrupoles alternate from site to site. However, while all 4$f$ multipolar models associated with the [1/2~1/2~1/2] ordering wave-vector should break the cubic symmetry (see an example, left, on figure \ref{Fig5}), no symmetry lowering is detected in CeB$_6$ \cite{Amara2012}.\\
To clarify this puzzle, one should reconsider the paramagnetic state. As regards the CEF, experiments indicate that the degeneracy of the $\Gamma_8$ quadruplet, essential to the quadrupolar ordering, is already reduced at temperatures much higher than T$_Q$. Inelastic neutron and Raman scattering experiments \cite{Zirngiebl1984, Loewenhaupt1985} show that, below 20 K, the ground state splits over an energy range of about 30 K. This is consistent with the temperature dependence of the magnetic entropy, obtained from specific heat measurements \cite{Fujita1980, PEYSSON1986}, which smoothly evolves above T$_Q$ and doesn't reach the quadruplet value $R\,ln4$ until temperatures higher than 30 K. At T$_Q$, the magnetic entropy value is close to $R\,ln(3.3)$.  As a J = 5/2 multiplet in $O_h$ point symmetry splits into a $\Gamma_7$ doublet and a $\Gamma_8$ quadruplet, one has few options but to :\\
- reverse the analysis of the spectroscopic experiments, adopting the $\Gamma_7$ level as a ground state. One will be then faced with many discrepancies, notably the temperature dependence of the width and position of the CEF excitation.\\
- accept that, as the temperature is reduced, the symmetry of the Ce$^{3+}$ environment is no longer cubic. As there is no track, from the crystallographic analysis, of a change affecting the Ce$^{3+}$ site above T$_Q$, the $O_h$ symmetry is preserved on average.\\
The latter option is consistent with a centrifugal Jahn-Teller effect. While the Jahn-Teller effect increases the out-of-centre probability of presence of the RE ion, it doesn't change the symmetry of its distribution. Furthermore, the effect is temperature dependent : at high enough temperature, all CEF states within the displacement split electronic ground state are equally populated and the centrifugal effect cancels. As a result, the magnetic entropy can evolve on a temperature scale, related to $W$, much smaller than the $\Gamma_8$ - $\Gamma_7$ difference.\\
The above Jahn-Teller scenario will obviously have consequences on the ordering properties of CeB$_6$. With a Ce$^{3+}$ already excentred above T$_Q$, pair couplings can easily drive an ordering of cage multipoles. Looking at a cage multipolar solutions that could solve the symmetry puzzle, i.e. an ordering with [1/2~1/2~1/2] wave vector that preserves the cubic symmetry, the lowest satisfactory solution consists in an order of the $\Gamma_1^-$ octupoles (right on figure \ref{Fig5}). However, our simplified spherical model, restricted to the consideration of the 1$s$ and 2$p$ levels, results in a null $\Gamma_1^-$ octupolar susceptibility and cannot account for such an octupolar ordering. The cage potential in CeB$_6$ is presumably strongly anisotropic, as is the crystal field in this material, and can mix different spherical vibration states, beyond the 2$p$ level. Then, a non-zero $\Gamma_1^-$ octupolar susceptibility could be expected.

\section*{References}
\bibliography{/Applications/TeX/biblio.bib}

\end{document}